\documentstyle[aps,prc]{revtex}
\begin{document}
\draft
\title{
Applicability of the orientation average formula in heavy-ion fusion reactions 
of deformed nuclei
}
\author{Tamanna Rumin$^{1}$,
\thanks{Electronic address: rumin@nucl.phys.tohoku.ac.jp}
Kouichi Hagino$^{2}$, 
\thanks{Electronic address: hagino@phys.washington.edu}
and
Noboru Takigawa$^{1}$
\thanks{Electronic address: takigawa@nucl.phys.tohoku.ac.jp}
}
\address{$^1$ Department of Physics, Tohoku University, 
Sendai 980-8578, Japan \\
$^2$ Institute for Nuclear Theory, Department of Physics,
University of Washington, Seattle, Washington 98195
}
\date{\today}

\maketitle

\begin{abstract}

In heavy-ion fusion reactions involving a well deformed nucleus, 
one often assumes that the orientation of the target nucleus does 
not change during the reaction. We discuss 
the accuracy of this procedure by 
analyzing the excitation function 
of the fusion cross section and the fusion barrier distribution in 
the reactions of $^{154}$Sm target with 
various projectiles ranging from $^{12}$C to $^{40}$Ar. 
It is shown that the 
approximation gradually looses its accuracy with increasing 
charge product of the projectile and target nuclei because of 
the effects of finite excitation energy of the target nucleus. 
The relevance of such inaccuracy 
in analyzing the experimental data is also discussed.

\end{abstract}

\pacs{
25.60.Pj,     
21.60.Ev,     
24.10.Eq,     
25.           
}

It is now well established that nuclear intrinsic degrees of 
freedom strongly influence the fusion cross section in heavy-ion 
reactions at energies near and below the Coulomb barrier \cite{bt98,DHRS98}. 
Typical examples include the rotational excitation of a deformed target 
nucleus which leads to a large enhancement of the fusion cross section 
at low energies. 
A characteristic in this case is that the excitation energy of the rotational 
motion is often much smaller than the curvature of the fusion barrier, which 
determines the time scale of the fusion process. This is the case when, e.g., 
one of the deformed rare earth nuclei or actinides is the target nucleus. 
One then often calculates the fusion probability for each partial wave $J$
following \cite{nbt86,htbd95}
\begin{eqnarray}
P^S_J(E)=\int_0^{1} P_J(E,\theta)d(\cos\theta) 
\label{orient}
\end{eqnarray} 
where $\theta $ is the orientation of the deformed target, which will 
be specified later more precisely. 
We call this the orientation average formula. 
The upper index $S$ stands for the 
sudden tunneling approximation. $P_J(E,\theta)$ is the fusion 
probability for a given orientation. It is determined by solving a 
one-dimensional Schr\"odinger equation for the relative distance 
between the projectile and target $R$, where 
the potential consists of the nuclear and Coulomb components given by 
\begin{eqnarray}
V_N(R,\theta)&=&\frac{-V_0}{1+\exp[(R-R_P-R(\theta))/a]}
\label{nucl}
\\
V_C(R,\theta)&=&\frac{Z_PZ_Te^2}{R}+\sum_{\lambda}
\left[\beta_{\lambda}+
\frac{2}{7}\sqrt\frac{5}{\pi}
\beta_{\lambda}^2\,\delta_{\lambda,2}\right]
\frac{3Z_PZ_Te^2}{2\lambda+1}\frac{R_T^\lambda}{R^{\lambda+1}}
Y_{\lambda 0}(\theta)
\label{coul}
\end{eqnarray}
with the angle dependent radius of the target nucleus
\begin{eqnarray}
R(\theta)=R_{T}\Big[1+\sum_{\lambda}\beta_{\lambda}Y_{\lambda 0}
(\theta)\Big].
\label{radius}
\end{eqnarray}
In these equations $V_0$ is the depth parameter of the nuclear potential 
between the projectile and target, $R_P$ ($R_T$) and $Z_P$ ($Z_T$) are the 
radius and the atomic number of the projectile (target), $\beta_\lambda $ is 
the deformation parameter of the target nucleus of multipolarity $\lambda $, 
and $a$ the surface diffuseness parameter. 
The second order term is retained only for the quadrupole 
coupling in the Coulomb interaction. 
As in almost all analyses of 
heavy-ion fusion reactions at sub-barrier energies, we adopt the no-Coriolis 
approximation and work in the rotating coordinate frame 
\cite{ti86,tab92,htbb95}, 
where the z-axis is taken to be parallel to the radial vector of the 
relative motion between the projectile and the target nuclei. 
Also, we assume an axial symmetry for the target nucleus. 
The parameter $\theta $ is then the angle between the 
symmetry axis and the z-axis in the rotating frame. 
 
Eq. (\ref{orient}) is quite appealing and 
provides a simple understanding of the 
large enhancement of the fusion cross section at sub-barrier energies. 
It states that the deformation and the associated rotational excitation of the 
target nucleus result in a distributed fusion barriers, 
where some of them will be 
lower than that of the bare potential in 
the absence of these degrees of freedom. 
The same idea of the orientation average is still 
correct if one replaces the Gauss integral in Eq. (\ref{orient}) by an 
appropriate summation even when the rotational band is truncated at 
a certain angular momentum $I_{max}$ as is always the case in nuclear physics.
If only $\lambda=2$ is present, the corresponding sum is given by 
the $I_{max}+2$ points Gauss quadrature \cite{nbt86}. 

The orientation average procedure is exact only in the 
degenerate spectrum limit, i.e., 
when the excitation energy of the rotational motion is zero, which 
is not the case in actual nuclei. 
It is, therefore, important to examine the applicability of this formula 
before one makes a quantitative analysis of the experimental data. 
A step towards this direction has been undertaken in Ref. \cite{htbd95}, 
where an analytic formula has been derived 
to modify the fusion probability in the 
sudden tunneling approximation by 
taking the finite excitation energy of the rotational 
motion into account as 
\begin{eqnarray}
P_J(E)\approx \exp\left[-\frac{4}{15}\left(\frac{\sqrt\frac{5}{4\pi}
\beta_2 F(R_B) T_0}
{\hbar}\right)^2 \frac{\epsilon^*_{2^+}T_0}{\hbar}\right]P^S_J(E)
\label{analy}
\end{eqnarray}  
where $\epsilon^*_{2^+}$ is the excitation energy of the first excited 
2$^+$ state of the rotational band. $T_0=\pi/\Omega$, 
$\Omega$ being the curvature of the fusion barrier, is the tunneling time 
and $F$ is a measure of the strength of the channel coupling 
which causes the rotational excitation of the target nucleus during the 
collision. In Ref. \cite{htbd95}, it was identified with the coupling 
strength at the fusion barrier $R_B$. 

Eq. (\ref{analy}) is useful to qualitatively discuss 
the conditions to apply the 
orientation average formula. It clearly shows that 
the excitation energy of the rotational motion should be 
much smaller than the curvature of the fusion barrier. 
It also shows that the coupling strength also governs the 
validity of the formula. 
Despite these advantages, as we see later, Eq. (\ref{analy})
is not tolerable for quantitative discussions.  
This is partly because Eq. (\ref{analy}) has been derived in 
a perturbation theory, and also by ignoring the radial dependence of the 
coupling form factor. In heavy-ion fusion reactions, the latter 
is a very crude approximation, because the nuclear and Coulomb couplings 
interfere leading to a strong radial dependence of the coupling form 
factor\cite{HTB97}. 
Nevertheless, for future reference, we show in Table I 
the exponential factor 
in Eq. (\ref{analy}), which we call the dissipation factor, 
for several systems to be examined later. 
Table II lists the bare nuclear potential parameters we used and the Coulomb 
barrier properties, where the radius parameter 
$r_0$ has been introduced as $R_P+R_T=r_0(A_P^{1/3}+A_T^{1/3})$.  

In this paper we compare the results of the orientation average 
formula with the numerical solutions of the corresponding coupled channels 
equations, which are obtained using the 
computer code CCFULL\cite{HRT99} by keeping 
the finite excitation energy, i.e. 0.082 MeV, 
of the target nucleus $^{154}$Sm. We call the latter as the 
exact coupled channels calculations. 
We keep only the quadrupole deformation in the target, i.e., 
$\lambda=2$, for simplicity.
All the projectiles are treated as inert. 
We truncate the rotational band of $^{154}$Sm at $I_{max}$=20$^{+}$ member 
for all reactions. According to 
the table of isotopes \cite{toi}, the highest member 
which has been so far experimentally observed is 16$^+$. 
We have confirmed that the comparison 
between the exact coupled-channels 
and the orientation average calculations does not almost change 
beyond $I_{max}$=12$^+$. 
We also remark that 
the coupled-channels calculations almost converge at 
$I_{max}$=12$^{+}$ member\cite{tht00} for the $^{16}$O and $^{154}$Sm 
reactions, but higher levels introduce non-negligible effects 
for heavy projectiles. 

Both the orientation average and the exact coupled channels calculations 
give the fusion cross section which is a monotonically 
increasing function of energy. 
In order to facilitate to see the accuracy of the 
orientation average formula, we plot in Fig. 1 the ratio of the 
fusion cross section calculated by solving the exact coupled channels 
equations to that obtained by the orientation average formula. 
The figure clearly shows that the deviation of the results of the orientation 
average formula from those of the exact calculations gradually increases 
with the charge product of the projectile and target nuclei.  
This behaviour 
is consistent with Table I, which has been obtained based on 
Eq. (\ref{analy}), though 
the actual deviation is much larger than what Eq. (\ref{analy}) predicts.
We also note that the deviation is 
significant even for light projectiles at energies below 
the Coulomb barrier. 

An important question is whether 
this significantly affects 
the understanding of the mechanism of heavy-ion fusion reactions and 
the structural informations such as the deformation parameters 
extracted from the data analyses. In this connection, we compare 
in Fig. 2 the experimental data of the excitation function 
of the fusion cross section and the fusion barrier distribution 
\cite{RSS91,LDH95} 
with the theoretical results calculated by the orientation average 
formula (the dashed line) and by the exact treatment of the corresponding 
coupled-channels equations (the solid line) for $^{16}$O+$^{154}$Sm reaction. 
The second derivative of the 
cross section times the bombarding energy has been calculated 
with the point difference method with the interval $\Delta E=2$ MeV. 
Similar comparison is done in Fig. 3 for $^{40}$Ar+$^{154}$Sm reaction.
The bare Coulomb barrier is at $V_B=59.41$ MeV and 127.57 MeV 
for $^{16}$O+$^{154}$Sm and $^{40}$Ar+$^{154}$Sm reactions, respectively.
In these calculations the deformation parameter 
$\beta_2$ of $^{154}$Sm was determined to be 0.32 by fitting 
the data of $^{16}$O+$^{154}$Sm fusion reactions. The 
same deformation parameter has been used for $^{40}$Ar+$^{154}$Sm reactions 
as well, 
though the effective optimum values can differ in two reactions. 
Fig. 2 shows that the difference between two theoretical lines 
is much smaller than their deviation from the experimental data. 
This indicates that one can safely use the orientation average 
formula to data analyses of $^{16}$O+$^{154}$Sm fusion reactions, 
even though its deviation from the 
exact calculation can be noticeable at low energies as shown in Fig. 1. 
The top panel of Fig. 3 shows that the deviation between the exact and 
the orientation average calculations is not so drastic even 
for $^{40}$Ar+$^{154}$Sm reaction in this semilogarithmic plot. 
On the other hand, the situation is different for the 
fusion barrier distribution shown in the lower panel.  
The difference between the two calculations can be more easily 
recognizable than the case for $^{16}$O+$^{154}$Sm fusion reactions. 
The present theoretical calculations still underestimate the 
fusion cross sections compared with the experimental data
for the $^{40}$Ar+$^{154}$Sm reaction at low energies. 
This will be partly because we ignored the projectile excitations, 
whose excitation energy is as small as 1.46 MeV. 

In summary, we have studied the accuracy of the orientation average 
formula for fusion cross sections between a spherical 
projectile and a well deformed target by comparing its results 
with those of the exact coupled channels calculations. 
We found that the results of the orientation average 
formula significantly differ from those of the exact numerical solution 
of the corresponding coupled channels equations  
for systems with a large charge product of the projectile 
and target nuclei. 
This suggests the necessity of the proper coupled-channels 
calculations beyond the orientation average formula 
for these systems 
in order to properly identify the role of various channel coupling effects and 
to extract reliable informations on nuclear structure, especially 
through the analysis of the fusion barrier distribution of 
high precision data. Unexpectedly, we observed a significant 
deviation of the orientation average formula from the exact calculations 
even for light projectiles in the energy region well below the 
Coulomb barrier. This deviation is, however, much smaller than 
the deviation of these two calculations from the experimental data. 
In this sense, the orientation average formula is 
safely applicable to light systems. 

\section*{Acknowledgments}

We thank D.M. Brink for useful discussions. 
This research was supported by the Monbusho Scholarship and 
the International Scientific Research Program: Joint Research: contract 
number 09044051 from the Japanese Ministry of Education, Science and Culture, 
and by the U.S. Department of Energy under Grant no. DE-FG03-00-ER41132.


\newpage

\medskip

TABLE I : Dependence of the dissipation factor 
on the projectile in the fusion of 
$^{154}$Sm target estimated based on Eq. (5). 

\begin{center}
\begin{tabular}{|l|l|l|l|l|l|l|l|l|l|l|}
\hline 
Reactions & $^{12}$C & $^{16}$O & $^{20}$Ne & $^{24}$Mg & $^{28}$Si 
& $^{32}$S & $^{36}$Ar & $^{40}$Ar \\
\hline
$^{154}$Sm & 0.810 & 0.689 & 0.567 & 0.456 & 0.357 & 0.274 & 0.207 & 0.164 \\
\hline
\end{tabular}
\end{center} 

TABLE II : The bare nuclear potential parameters for different systems. 

\begin{center}
\begin{tabular}{|l|l|l|l|l|l|l|l|l|l|l|}
\hline 
Systems & $V_0$ (MeV)& $r_0$ (fm) & $a$ (fm) & $V_B$ (MeV) 
& $R_B$ (fm) & $\hbar\Omega$ (MeV) \\
\hline
$~{12}$C+$^{154}$Sm & 150.0 & 0.950 & 1.05 & 44.60 & 10.81 & 3.47 \\
\hline
$^{16}$O+$^{154}$Sm & 165.0 & 0.950 & 1.05 & 59.41 & 10.81 & 3.48 \\
\hline
$^{20}$Ne+$^{154}$Sm & 190.0 & 0.945 & 1.05 & 73.85 & 10.87 & 3.48 \\
\hline
$^{24}$Mg+$^{154}$Sm & 225.0 & 0.935 & 1.05 & 88.04 & 10.95 & 3.50 \\
\hline
$^{28}$Si+$^{154}$Sm & 255.0 & 0.935 & 1.05 & 101.58 & 11.09 & 3.50 \\
\hline
$^{32}$S+$^{154}$Sm & 285.0 & 0.935 & 1.05 & 114.89 & 11.22 & 3.51 \\
\hline
$^{36}$Ar+$^{154}$Sm & 315.0 & 0.935 & 1.05 & 128.00 & 11.34 & 3.51 \\
\hline
$^{40}$Ar+$^{154}$Sm & 294.0 & 0.935 & 1.05 & 127.57 & 11.38 & 3.34 \\
\hline
\end{tabular}
\end{center} 

\newpage


\begin{center}
{\Large Figure Captions}
\end{center}

\noindent

{\large FIG. 1}\\
The ratio of the fusion cross sections obtained in the  
exact coupled channels calculations 
to those obtained in the orientation average formula. 

\noindent 

{\large FIG. 2}\\
Comparison of the experimental data and the results of the orientation 
average (the dashed line) and the exact coupled channels 
(the solid line) calculations for the 
$^{16}$O+$^{154}$Sm fusion reaction. The top and bottom panels are 
the fusion excitation function and the fusion barrier distribution,
respectively. The data are taken from Ref. \cite{LDH95}.

\noindent

{\large FIG. 3}\\
The same as Fig. 2, but for the $^{40}$Ar+$^{154}$Sm fusion reaction. 
The data are taken from Ref. \cite{RHH85}.

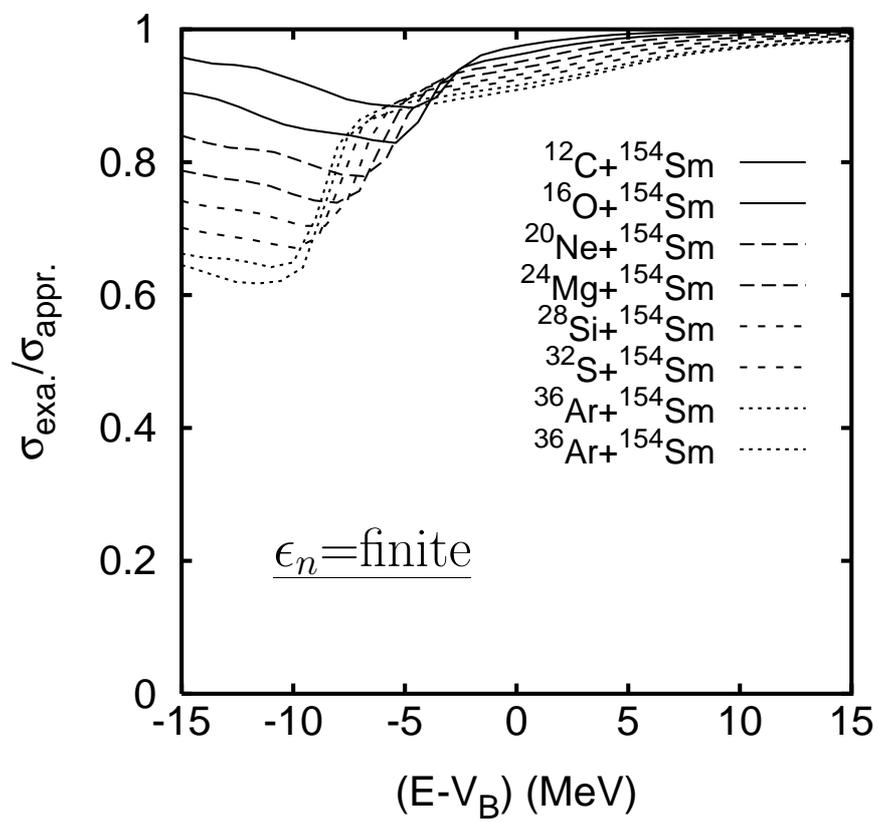
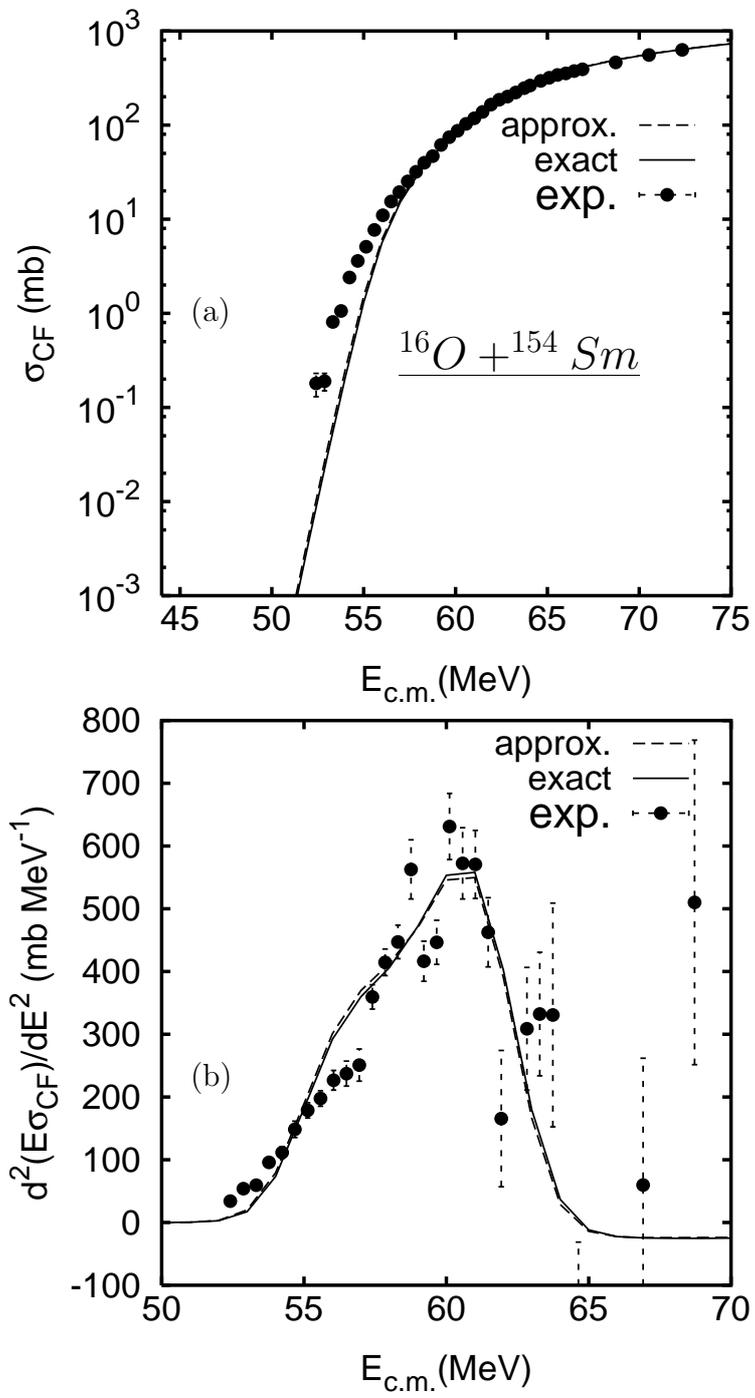
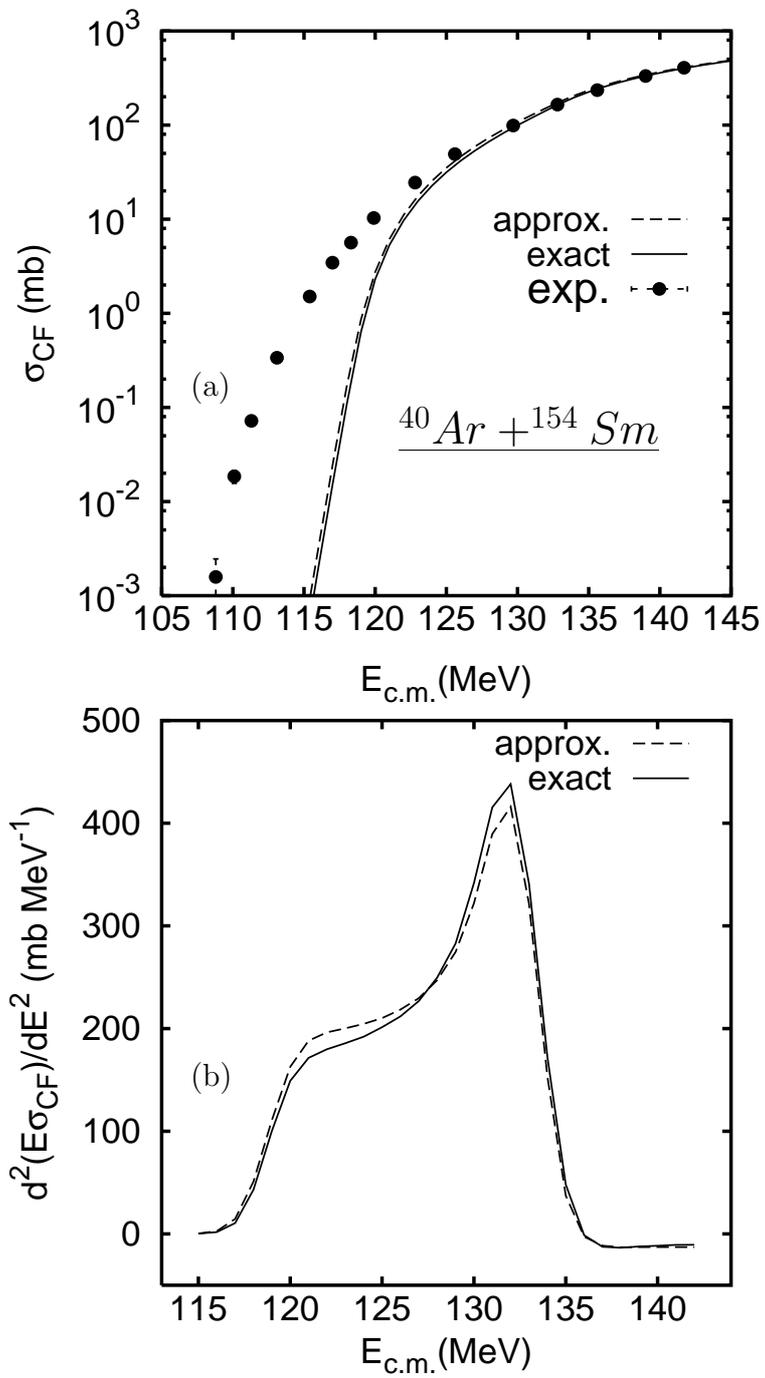
\end{document}